# Increased pump acceptance bandwidth in spontaneous parametric downconversion process using Bragg reflection waveguides

**Krishna Thyagarajan[1], Ritwick Das[1], Olivier Alibart[2], Marc de Micheli[2], Daniel B. Ostrowsky[2], and Sébastien Tanzilli[2]**

[1]*Department of Physics, Indian Institute of Technology Delhi, New Delhi 110016, India*
[2]*Laboratoire de Physique de la Matière Condensée, CNRS UMR 6622, Université de Nice Sophia-Antipolis, Parc Valrose, 06108 Nice Cedex 2, France*
ktrajan@physics.iitd.ac.in, ritwick.das@gmail.com

**Abstract:** In this paper we show that by suitably tailoring the dispersion characteristics of a Bragg reflection waveguide (BRW) mode, it is possible to achieve efficient photon pair generation over a large pump bandwidth while maintaining narrow signal bandwidth. The structure proposed consists of a high index core BRW with a periodically poled GaN core and periodically stratified cladding made up of alternate layers of $Al_{0.02}Ga_{0.98}N$ and $Al_{0.45}Ga_{0.55}N$. Such photon-pair generators should find applications in realizing compact and stable sources for quantum information processing.

## 1. Introduction

Efficient generation of entangled photon pairs using spontaneous parametric downconversion (SPDC) have been the key ingredient for a number of quantum optics experiments, ranging from testing the foundations of quantum mechanics to various quantum communication applications such as quantum cryptography [1] and quantum teleportation [2-4] over long distances. The use of periodically poled waveguides as SPDC sources has shown a great potential for enhancing the flux of correlated photon pairs generated with relatively small pump powers due to long effective interaction lengths [5-9]. The photon pair generation efficiency of a waveguide based SPDC source essentially depends on the nonlinearity of the constituent materials, modal overlap of the interacting waves, and the interaction length [6,10]. However, most of the above mentioned quantum experiments take advantage, for accurate timing purpose, of ultra-short laser pulses, which implies a pump bandwidth on the order of several nanometers. In addition, the downconverters used in these experiments are designed, together with additional filters, for specific pairs of signal and idler wavelength windows that set, through to the energy conservation law, a central pump wavelength for which the conversion efficiency is maximal. Now, if the pump deviates from this central wavelength, the downconversion efficiency at the designed signal wavelength is reduced. We therefore define the pump acceptance bandwidth (BW) of any downconverter as the spectral width over which the conversion efficiency remains higher than half its maximum value at a chosen signal wavelength. This pump acceptance BW is a consequence of the so-called phase-matching condition. Hence, if we have a pump that has a spectral width of about the pump acceptance BW, then a majority of the input pump photons would be utilized in the downconversion process. On the other hand, increasing for instance the length of a non-linear waveguide would increase the conversion efficiency but the signal BW would still be substantially broad due to near identical dispersion characteristics of the signal and idler modes. In the latter case, using a broadband pump would usually result in a very broad spectrum of the generated signal photons which is inescapably submitted to fiber chromatic dispersion as reported in Refs. [7,11]. Usually the signal photon within a narrowband region is filtered out to reduce the adverse impact of fiber dispersion for long distance propagation but this inevitably leads to reduction in efficiency [7,9,11]. One intuitive route which is usually followed to obtain narrow signal BW is to choose the signal and idler modes to be orthogonally polarized (type-I or type-II interaction) that exhibit substantially different dispersion characteristics [8,12]. Recently, Fujii *et al.* [8] have experimentally demonstrated the generation of narrowband signal photons (~ 1 nm BW for 30 mm long waveguide) using type-II interaction in a periodically poled lithium niobate waveguide.

In the present work, we show that the use of Bragg reflection waveguides (BRWs) allows achieving narrow signal BW together with broad pump acceptance BW using a type-0 interaction with high efficiency in a $GaN/Al_xGa_{1-x}N$ based nonlinear system. For quite sometime now, GaN has been investigated as an efficient nonlinear material due to its wide transparency range (365 nm–13.6 μm) and high nonlinear coefficient ($d_{33}$ = 16.5 pm/V), comparable to LiNbO$_3$ [13,14]. The use of epitaxially grown III-V semiconductors like GaN (and their ternary derivatives) for the design of optoelectronic devices is particularly advantageous because of the wide possibility of exploiting the mature fabrication technology for designing monolithically integrated components [14]. Exploiting the high second order nonlinearity of GaN, Chowdhury *et al.* [15] have experimentally demonstrated SHG in periodically poled GaN (PPGaN) but the nonlinear properties of GaN were not fully exploited due to the strong dispersion exhibited by bulk GaN in the visible region.

The BRWs are a class of interferometric waveguides with high or low index core and periodically stratified cladding which act as Bragg reflectors confining energy in the core as shown in Fig. 1 [16]. The strong phase velocity dispersion characteristics exhibited by the BRWs [17] can be suitably exploited to obtain the desired dispersion for pump, signal or idler modes which allows controlling the efficiency via the phase matching condition. Note that the modal phase velocity dispersion is very strong in the visible region essentially due to strong material dispersion exhibited by most of the ionic and semiconducting nonlinear materials whereas in the mid-IR range (close to 1550 nm band) the dispersion is relatively weak [15,18]. This is the reason why a narrowband pump at 800 nm is always downconverted into broadband signal and idler photons at 1550 nm. The central idea behind this work is to entail appropriate dispersion in the idler mode using the waveguide geometry so that it counters the strong dispersion of the pump mode so as to increase the pump acceptance bandwidth and at the same time destroys the phase matching condition rapidly if the signal wavelength deviates from the design wavelength.

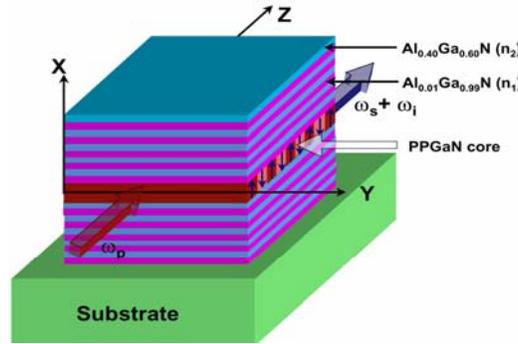

Fig. 1. Schematic of the proposed high index core BRW design for SPDC with PPGaN core and periodic cladding of Al$_{0.02}$Ga$_{0.98}$N ($n_1$) and Al$_{0.45}$Ga$_{0.55}$N ($n_2$) layers. The thicknesses of the layers corresponding to refractive indices $n_c$, $n_1$, $n_2$ are labeled $d_c$, $d_1$, $d_2$ respectively.

## 2. Results and analysis

We consider a high index core symmetric planar BRW with PPGaN core [14] and periodically stratified cladding made up of alternate layers of Al$_{0.02}$Ga$_{0.98}$N ($n_1$) and Al$_{0.45}$Ga$_{0.55}$N ($n_2$). These two particular alloys provide significant contrast in the refractive indices for better confinement of the modes as well as assisting in obtaining suitable dispersion characteristics [17]. The epitaxial growth technique could lead to periodic poling of the upper cladding layers as well but this would have no effect on the argument presented here. The pump ($\lambda_p$ = 800 nm) and signal ($\lambda_s$ = 1550 nm) are taken to be total internal reflection (TIR) guided modes (*x*-polarized) in the high index core BRW (Fig. 1) and the idler ($\lambda_i$ = 1653 nm) is taken to be a BRW mode (*x*- polarized) that satisfies the quarter wave condition [17]. The analysis of the BRW modes is carried out by considering only the modes which have effective indices lower than those of the BRW materials ($n_c$, $n_1$ and $n_2$), and falling within the stop band of the periodic cladding [16]. The calculation of the effective indices for the BRW modes is described in [16,17]. Since, the guidance takes place via Bragg reflection; the BRW modes are highly dispersive essentially due to strong variation of the reflection coefficient of the periodic cladding with the wavelength [17]. The TIR guided modes are analyzed using the standard transfer matrix method for leaky waves with 12 bilayers of the periodic cladding to achieve negligible leakage loss [18]. The refractive index and nonlinear coefficients of various components of the waveguiding structure is calculated using references [19] and [13], respectively. The proposed structure is designed such that the phase velocity dispersion slope of the idler (BRW) mode almost matches with that of the pump whereas it is significantly different from that of the signal. This can simultaneously lead to broad pump acceptance BW and narrow signal BW.

The quasi phase matching (QPM) condition is given by,

$$\beta_p = \beta_s + \beta_i + K \qquad \ldots (1)$$

where $\beta_p$, $\beta_s$ and $\beta_i$ are the propagation constants of the mode at the pump, signal and idler wavelengths respectively and $K$ represents the grating vector of the QPM grating. Now, for a fixed grating period, the change in signal wavelength due to a change in pump wavelength is given by,

$$\frac{d\beta_s}{d\lambda_p} = \frac{d}{d\lambda_p}(\beta_p - \beta_i) \qquad \ldots (2)$$

If the BRW is suitably designed then at the chosen $\lambda_p$, the difference between the propagation constants ($\beta_p - \beta_i$) of the pump and the idler modes could exhibit a minimum variation with respect to $\lambda_p$ [20]; this would imply that as $\lambda_p$ changes, the signal wavelength ($\lambda_s$) would remain almost fixed, thus resulting in a large pump acceptance BW. For this to happen we need to design the BRW structure so that the strong modal dispersion (essentially due to material dispersion) at $\lambda_p$ is countered by strong waveguide dispersion at the idler wavelength ($\lambda_i$). Figure 2 plots the variation in ($\beta_p - \beta_i$) with $\lambda_p$ for a BRW with a core thickness $d_c = 582$ nm; the thicknesses of the cladding layers are chosen to be $d_1 = 293$ nm and $d_2 = 517$ nm so as to satisfy the quarter wave stack condition exactly [17]. For $\lambda_p = 800$ nm and $\lambda_s = 1550$ nm, using these waveguide parameters, the spatial period of the nonlinear grating for QPM is $\Lambda_{QPM} = 2.77$ μm. As shown in Fig. 2 the designed structure shows a minimum at the chosen pump wavelength of 800 nm, thus giving the possibility of a large pump acceptance BW.

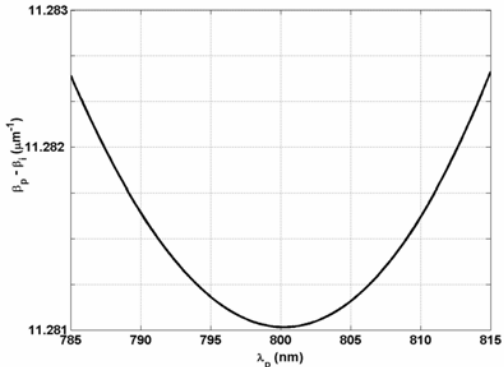
Fig. 2. Plot showing the variation of the difference of the propagation constants of the pump (TIR-guided) and idler (BRW) modes, ($\beta_p$-$\beta_i$) as a function of the pump wavelength ($\lambda_p$) for $d_c = 582$ nm and $\Lambda_{QPM} = 2.77$ μm.

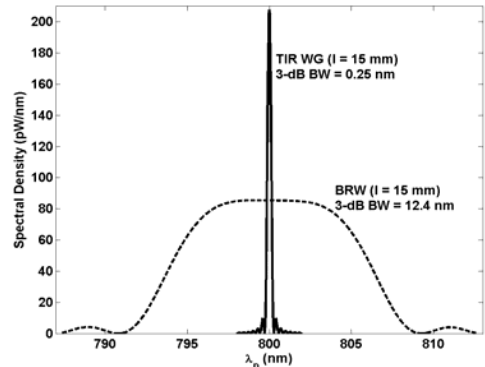
Fig. 3. Plot of the signal spectral power density as a function of pump wavelength ($\lambda_p$) with $\lambda_s = 1550$ nm for SPDC in the BRW structure (dashed curve) and in the conventional TIR based waveguide (solid curve).

In order to illustrate the idea, we have calculated the signal spectral power density variation with respect to pump wavelength ($\lambda_p$) for SPDC in the waveguide using the following equation [12],

$$\frac{dP_s}{d\lambda_s} = \frac{16\pi^3 \hbar d_{eff}^2 l^2 c P_p}{\varepsilon_0 n_s n_p n_i \lambda_s^4 \lambda_i} I_{ov}^2 sinc^2\left(\frac{\Delta\beta l}{2}\right) \qquad \ldots (3)$$

where $d_{eff}$ (= $(2/\pi)d_{33}$) is the effective nonlinear coefficient for first order QPM, $\Delta\beta = \beta_p - \beta_s - \beta_i - 2\pi/\Lambda_{QPM}$ is the phase mismatch parameter. '$l$' is the interaction (waveguide) length, $P_p$ is the input pump power per unit length in the y-direction (invariant) and $n_s$, $n_p$ and $n_i$ are the effective indices of signal ($\lambda_s$), pump ($\lambda_p$) and idler ($\lambda_i$) modes respectively. '$I_{ov}$' is the overlap integral given by [12],

$$I_{ov} = \int dx E_x^{(p)} E_x^{(s)} E_x^{(i)} \qquad \ldots (4)$$

where the modes have been normalized according to $\int dx |E_x^{(p,s,i)}|^2 = 1$. Keeping the signal wavelength fixed at 1550 nm, Fig. 3 shows the signal spectral power density variation as a function of $\lambda_p$ showing a flat region close to $\lambda_p$ = 800 nm with a 3 dB pump acceptance bandwidth $\Delta\lambda_{FWHM}$ = 12.4 nm for an interaction length of $l$ = 15 mm and input pump power of 1 mW/μm under cw operation. For comparison, we have also plotted the signal spectral power density curve for a conventional symmetric planar waveguide with PPGaN core and $Al_{0.20}Ga_{0.80}N$ cladding for identical input pump power of 1 mW/μm and same length ($l$ = 15 mm). The thickness of PPGaN core is assumed to be 700 nm and the phase matching takes place between fundamental TIR guided modes at the pump, signal and idler. The BW obtained in such a conventional geometry is $\Delta\lambda_{FWHM}$ = 0.25 nm. The unavoidable tradeoff for the broad pump acceptance BW in a suitably designed BRW is the fall in peak signal spectral power density value as compared to the conventional geometry. This is mainly because of the reduction in modal overlap in Eq. (4) due to oscillatory behaviour of the idler (BRW) mode in the periodically stratified cladding [16,17]. The large pump spectral BW in our proposed waveguide structure is essentially obtained by an appropriate design in which variations in $\lambda_p$ are absorbed by the idler keeping $\lambda_s$ almost fixed. To make this point clear, in Fig. 4 we have plotted the variation of the phase-matched values of $\lambda_s$ and $\lambda_i$ as a function of $\lambda_p$ for the designed BRW. It can be seen that $\lambda_s$ changes insignificantly over the entire pump wavelength range of interest (793-806 nm) whereas $\lambda_i$ curve exhibits a positive slope in accordance with the requirement of conservation of energy for the three-wave interaction process. Thus this interaction would lead to a very narrow bandwidth for the signal and a large bandwidth for the idler. Narrow bandwidth signal photon generation can be very useful for SPDC using broadband pumps or ultrashort pulses of few hundred-femtosecond regimes.

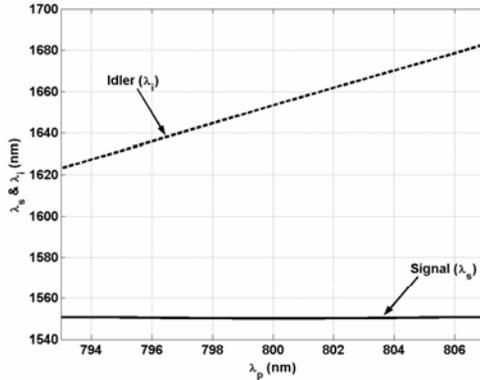

Fig. 4. Plot of variation of signal ($\lambda_s$) and idler ($\lambda_i$) wavelengths as function of pump wavelength ($\lambda_p$) for the BRW with $d_c$ = 582 nm and $\Lambda_{QPM}$ = 2.77 μm.

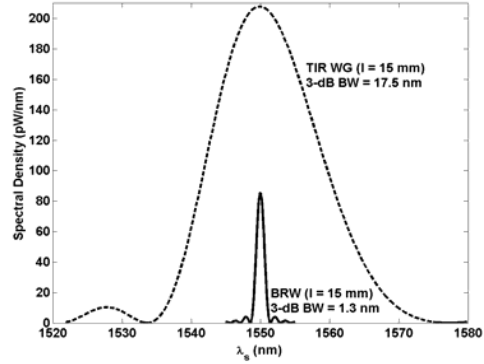

Fig. 5. Plot of the signal spectral power density as a function of signal wavelength ($\lambda_s$) with $\lambda_p$ = 800 nm for SPDC in the BRW structure (solid curve) and in the conventional TIR based waveguide (dashed curve).

Keeping now the pump wavelength fixed at 800 nm, Fig. 5 shows the variation of signal spectral power density as a function of $\lambda_s$, for the designed BRW as well as the conventional

TIR waveguide showing a signal BW of $\Delta\lambda_{FWHM}$ = 1.3 nm for $l$ = 15 mm, which is much smaller for the designed BRW as compared to the conventional case ($\Delta\lambda_{FWHM}$ = 17.5 nm) for $l$ = 15 mm. The reason for this narrow BW is that the dispersion characteristics of the signal (TIR guided) and idler (BRW mode) are so different that considerable phase mismatch ($\Delta\beta \neq 0$) between the interacting modes results if $\lambda_s$ changes (with $\lambda_p$ remaining constant).

The expected number of photon pairs generated by the BRW as well as the conventional waveguide can be estimated by using Eq. (3). If we limit our detection window to 1 nm around central signal wavelength ($\lambda_s$ = 1550 nm), we obtain the estimated pair flux to be ~$6.67\times10^8$ pairs/s per mW/µm of pump for 15 mm long BRW whereas that for conventional waveguide, the estimated photon flux would be ~$1.62\times10^9$ pairs/s per mW/µm of pump for identical length. Although, the estimated photon flux generation rate appears to be higher for a pump wavelength $\lambda_p$ of 800 nm (corresponding to phase matched operation) for the conventional waveguide, the distinct advantage offered by our design is the tolerance to pump wavelength changes. Thus if the pump wavelength deviates from the central pump wavelength ($\lambda_p$ = 800 nm) even by 0.5 nm, the efficiency of the photon flux generation for the conventional waveguide drops to almost zero (as shown in Fig. 3) while for the proposed design even for a deviation of pump wavelength by ~ 6 nm, the photon pair generation efficiency would still be appreciably large with a corresponding generation of a narrowband signal (~ 1 nm).

### 3. Discussion

The primary advantage offered by BRWs for efficient generation of entangled photon pairs is that we can tune the dispersive behaviour of a BRW mode to obtain suitable BW with respect to any wavelength of operation and for any material system within the constraints of transparency and nonlinearity of the constituent materials. This implies that depending upon the dispersion exhibited by the constituent materials and the BW targeted for the three-wave interaction, the BRW mode could be chosen appropriately and its waveguide dispersion could be tailored accordingly so as to counter the dispersion of other guided modes. By forcing the BRW mode to operate closer to the center of the bandgap, we can enhance the overlap in Eq. (4) but this leads to a trade-off between the target BW and spectral power density. Nevertheless, the freedom afforded by this design to accomplish type-0 interaction using any material system with high nonlinear coefficients can be exploited to achieve high spectral power density. Even though, this idea has been recently exploited using bulk birefringent crystals [21], we believe the present work is the first proposition accomplishing this in a guided wave configuration that would lead to considerably higher efficiencies essentially due to longer interaction length and possibility of accessing the maximum nonlinear coefficient.

### 4. Conclusion

In conclusion, we have proposed a BRW design that can simultaneously achieve broad pump acceptance bandwidth and narrow signal bandwidth with only a small sacrifice of the efficiency. Due to the narrow signal BW, the idler spectrum would be broad if a broadband pump is used. Such a technique can be implemented for designing an unbalanced photon-pair source in terms of bandwidth, for which the narrowband photon would travel over long distance while the broadband one would be operated locally to prevent from decoherence due to dispersion. As an example, one could think of a heralded single photon source in which the broadband photon is detected, using an upconversion-based or a superconducting detector [22,23], in order to announce the arrival time of the associated narrowband signal photon at the end of the quantum communication channel [24].

### Acknowledgement

One of the authors (Ritwick Das) would like to thank the Council for Scientific and Industrial Research (CSIR), India for providing a Senior Research Fellowship. The work was partially supported by an Indo-French Networking project sponsored by DST, India and CNRS, France.